\documentstyle[prl,twocolumn,aps]{revtex}
\begin{document}
\draft
\preprint{HEP/123-qed}


\twocolumn[
\hsize\textwidth\columnwidth\hsize\csname
@twocolumnfalse\endcsname
\title{ Detailed study of the ac susceptibility of Sr$_2$RuO$_4$ in oriented magnetic fields}
\author{Hiroshi Yaguchi$^{1,2}$, Takashi Akima$^{1}$, Zhiqiang Mao$^{1,2}$\cite{maoaddress}, Yoshiteru Maeno$^{1,2}$, and Takehiko Ishiguro$^{1,2}$}
\address{$^1$Department of Physics, Graduate School of Science, Kyoto University, Kyoto 606-8502, Japan \\
$^2$ CREST, Japan Science and Technology Corporation, Kawaguchi, Saitama 332-0012, Japan }
\date{\today}

\maketitle

\begin{abstract}
\quad We have investigated the ac susceptibility of the spin triplet superconductor Sr$_2$RuO$_4$ as a function of magnetic field in various directions at temperatures down to 60 mK. We have focused on the in-plane field configuration (polar angle $\theta \simeq 90^{\circ}$), which is a prerequisite for inducing multiple superconducting phases in Sr$_2$RuO$_4$. We have found that the previous attribution of a pronounced feature in the ac susceptibility to the second superconducting transition itself is not in accord with recent measurements of the thermal conductivity or of the specific heat. We propose that the pronounced feature is a consequence of additional involvement of vortex pinning originating from the second superconducting transition.
\end{abstract}


]

\narrowtext
 
\section{Introduction}
\quad Sr$_2$RuO$_4$ is a layered perovskite superconductor without copper  \cite{discovery}. Despite its superconducting transition temperature $T_{\rm c}$ being rather low (ideally 1.5 K)  \cite{impurity,defect}, Sr$_2$RuO$_4$ has been of great interest because of its unconventional spin-triplet pairing. Soon after the discovery of its superconductivity, the possibility of spin-triplet pairing was pointed out on theoretical grounds \cite{Rice,Baskaran}. In fact, recent experiments have revealed its unconventional nature. In particular, the observation of spontaneous magnetic moments accompanying the superconducting state indicates broken time-reversal symmetry \cite{muSR}. Besides this, NMR measurements have demonstrated that the Knight shift is unaffected by the superconducting transition, providing a definitive indication of spin-triplet pairing with the spin of Cooper pairs lying within the ab-plane  \cite{NMR}.

\quad One of the most interesting aspects of spin-triplet superconductivity is that multiple superconducting phases could be induced owing to the Cooper-pairs possessing an internal degree of freedom. In fact, this is exemplified in UPt$_3$  \cite{UPt3}, which is another spin-triplet superconductor. Although the details of the superconducting wave function of Sr$_2$RuO$_4$ are still controversial  \cite{Cp,NQR,penetration,tanatar,izawa}, simple consideration based on existing experimental results will allow one to understand that the superconducting symmetry is probably represented by the degenerate two-component order parameter {\bf{\textit{d}}}(\textbf{\textit{k}}) = {\bf{\textit{z}}}$\Delta_{0}(k_x + ik_y)$  \cite{muSR,NMR,polarised,SANS}. Agterberg  \cite{theory} theoretically suggests that under such circumstances, this degeneracy in energy will be lifted in a magnetic field $H$ parallel to the ab-plane, leading to another superconducting phase with {\bf{\textit{d}}}(\textbf{\textit{k}}) = {\bf{\textit{z}}}$\Delta_{0}k_{x^\prime}$ being induced above a certain field $H_{2}$, where $H // x^\prime$. 
Agterberg also predicted that the appearance of this phase will be accompanied by an enhancement of the in-plane four-fold anisotropy of the upper critical field \cite{theory}.

\quad The occurrence of a second superconducting phase in Sr$_2$RuO$_4$ was first suggested by measurements of the ac susceptibility and the specific heat, which are reported in our previous paper  \cite{Mao}. We have attributed a clear kink in the ac susceptibility to the second superconducting transition and interpreted anomalous behaviour in the electronic specific heat as entropy release due to the second superconducting transition. The main conclusions of the previous \cite{Mao} paper may be summarised in the following way. The second superconducting transition occurs at a field slightly lower than the upper critical field $H_{\rm c2}$ only when the magnetic field is applied accurately parallel to the ab-plane. Concomitantly, the in-plane anisotropy of $H_{\rm c2}$ is significantly enhanced. Also a slight misalignment of the angle between the magnetic field and the ab-plane causes both the second superconducting transition and the enhancement of the in-plane anisotropy to be suppressed. Agterberg's theoretical prediction  \cite{theory} receives partial support from these experimental facts. Recent measurements of the specific heat  \cite{Cp} and of the thermal conductivity  \cite{kappa} have also detected a steep change attributable to the second superconducting transition, resulting in a field-temperature ($H$-$T$) phase diagram similar to the one deduced from the ac susceptibility measurements  \cite{Mao}.

\quad Nevertheless, there are significant discrepancies between theory and experiment in the $H$-$T$ phase diagram. For example, according to previous studies of the ac susceptibility  \cite{Mao}, the specific heat \cite{Cp} and the thermal conductivity  \cite{kappa}, the phase boundary between the two superconducting phases seems to merge with the upper critical field line at a bicritical point close to $H$ = 1.2 T, $T$ = 0.8 K. On the contrary, there should not be a bicritical point theoretically  \cite{theory}; the merging point is expected to be at $H = 0, T = T_{\rm c}$. 

\quad In this paper, we have extended to 60 mK the ac susceptibility measurements on Sr$_2$RuO$_4$ in magnetic fields along various directions, subsequent to the study by Mao {\it et al}.  \cite{Mao}; the measurements in ref. \cite{Mao} were carried out down to 0.35 K. We will compare in detail the signs of the second superconducting transition obtained by various experimental probes.

\section{Experiment}
\quad In our previous study of the ac susceptibility and the specific heat of Sr$_2$RuO$_4$  \cite{Mao}, we investigated three single crystals with different shapes chosen from different batches. In this paper, of those three samples, we concentrate on the sample referred to as sample B in ref. \cite{Mao} for detailed studies. The sample is a single crystal of Sr$_2$RuO$_4$ which was grown by a floating-zone method with an infrared image furnace  \cite{growth}. The sample was polished into a rectangle such that a side-surface of the sample and the a-axis make an angle of about $25^{\circ}$. The size was 1.9 $\times$1.3 $\times$ 0.5 ${\rm mm}^{3}$, with the shortest dimension along the c-axis. The sample was annealed in oxygen at 1050$^{\circ}$C for three weeks in order to reduce the amount of defect. A measurement of the ac susceptibility shows a sharp superconducting transition at $T_{\rm c}$ = 1.46 K (midpoint). An x-ray rocking curve of the sample shows the characteristics of a single crystal of high quality; the diffraction peak width [full width at half maximum (FWHM)] being comparable to that of a Si crystal (with FWHM of $0.06 ^{\circ}$) in the diffractometer. The directions of the tetragonal crystallographic axes of the sample were determined by x-ray Laue pictures.

\quad Low temperatures down to 60 mK were obtained by means of a dilution refrigerator. Temperatures were measured using a RuO$_2$ resistor. Magnetic fields of up to 2 T, generated by a superconducting solenoid, were applied to the sample.

\quad Measurements of the ac susceptibility were done by a mutual-inductance method with an alternating field of 50 $\mu$T at frequencies of 700-1000 Hz; the measurement frequencies were carefully chosen to ensure that frequency-dependent artefact was not involved. The ac modulation field was applied along the c-axis. The sample was mounted in a double-axis rotator that enables both the polar and azimuthal angles to be changed independently with a precision of $0.01 ^{\circ}$. (For referring to the direction of the applied magnetic field with respect to the crystallographic axes, we shall introduce the polar angle $\theta$, for which $\theta = 0^{\circ}$ corresponds to the [001] direction, and the azimuthal angle $\phi$, for which $\phi = 0^{\circ}$ with $\theta = 90^{\circ}$ corresponds to the [100] direction.) It is to be noted that the tetragonal symmetry of the crystal structure is conserved down to temperatures as low as 110 mK  \cite{tetragonal}.

\section{Results and Discussion}
\quad Figure 1(a) shows the ac susceptibility of Sr$_2$RuO$_4$ as a function of magnetic field parallel to the [110] direction, with an accuracy of $\Delta \theta \le 0.02^{\circ}$, at several temperatures. As in ref.  \cite{Mao}, three prominent features, labelled P$_{1}$, P$_{2}$ and P$_{3}$, are seen. Their definitions are illustrated in Figs. 1 (a) and (b); we follow the definitions used in our previous study \cite{Mao}. (i.e. We define $H_{\rm c2}$ as the intersection of the linear extrapolation of the most rapidly changing part of the real part of the ac susceptibility ($\chi = \chi^\prime + i\chi^{\prime\prime}$) and that of the normal state \cite{definition,demagnetisation}.) P$_{1}$ and P$_{3}$ correspond to the upper critical field $H_{\rm c2}$ and the second peak $H_{\rm p}$ due to vortex synchronisation  \cite{synchro}, respectively. These two features are seen at all the temperatures. On the other hand, the feature P$_{2}$, which was interpreted as a manifestation of a second superconducting transition  \cite{Mao}, appears only below $\sim$ 0.7 K. These observations confirm the results of Mao {\it et al}.  \cite{Mao} that cover temperatures down to 0.35 K; we have shown that the feature P$_{2}$ persists at least to 60 mK. Along the line of the discussion in the Agterberg theory \cite{theory}, the feature P$_{2}$ has been considered to represent a second-order transition from a state with a node-less gap to a state with lines of nodes (when the applied field is increased). In contrast, as shown in Fig. 1(b), when the applied field is along the [100], only the two features P$_{1}$ and P$_{3}$ are clearly seen, and P$_{2}$ is not observed in the ac susceptibility even at a temperature as low as 60 mK.  This is also consistent with our previous results  \cite{Mao}.

\quad In order to see how these features change in field position, the characteristic fields $H_{\rm c2}$, $H_{\rm 2}$ and $H_{\rm p}$ are plotted against the azimuthal angle $\phi$ (for $\theta = 90 ^{\circ}$, $T$ = 60 mK) in Fig. 2. All of the three features exhibit clear four-fold anisotropy, apart from that P$_{2}$ is not observed over $\sim 20 ^{\circ}$ around the [100] direction. The azimuthal-angle dependences of $H_{\rm 2}$ and $H_{\rm p}$ are both out-of-phase with that of $H_{\rm c2}$. $H_{\rm 2}$ appears to merge into $H_{\rm c2}$ and is not seen in the vicinity of $\phi$ = $0^{\circ}$. The in-plane anisotropy of the upper critical field is 3.5 \% at 60 mK.

\quad Figure 3 shows the $H$-$T$ phase diagrams for the $H //$ [100] case and for the $H //$ [110] case as contrasting examples. Whereas the Agterberg theory  \cite{theory} suggests that the second superconducting transition occurs in both cases, the feature P$_{2}$ disappears for $H //$ [100]. It should be noted that hysteresis was hardly observed between the upward and downward field sweeps; no obvious evidence for a first-order transition was seen.

\quad Contrary to the results of the ac susceptibility, a very recent study of the specific heat \cite{deguchi}, made subsequent to the work in ref  \cite{Cp}, has obtained definitive evidence for the second superconducting transition in magnetic fields along the [100] direction; they have observed a clear split at the superconducting transition in the temperature dependence of the specific heat. This observation shows good agreement with a recent study of the thermal conductivity  \cite{kappa}. Besides, the latter study suggests that the azimuthal angle dependence of $H_{\rm c2}$ and the second superconducting transition field show four-fold symmetry \cite{kappa}, but are in-phase with $H_{\rm c2}$, in apparent disagreement with the present ac susceptibility measurements. The clear split in the specific heat \cite{deguchi} provides {\it thermodynamic} evidence for a phase transition, and the transition point should not depend on the probe used. Therefore, these significant discrepancies inevitably cast doubt upon the attribution of P$_{2}$ seen in the ac susceptibility to the second superconducting transition itself.

\quad In connection to Fig. 3, it is worth mentioning that the temperature dependence of $H_{\rm p}$ at low temperatures below $\sim$ 0.4 K is unusual. $H_{\rm p}$ drops with decreasing temperature; the $H_{\rm p}$ line looks repelled by the $H_{\rm 2}$ line. This occurs immediately below the second superconducting phase in the $H$-$T$ phase diagram. (e.g. $H$-$T$ phase diagram shown in Fig. 8 of ref.  \cite{kappa}) A similar tendency can be seen in a close region of the phase diagram for $H //$ [100] despite the absence of P$_{2}$. Even though P$_{2}$ does not necessarily represent the second superconducting transition itself, the tendency suggests that there is a strong correlation between the second superconducting transition and the feature P$_{2}$. 

\quad Also the thermal conductivity  \cite{kappa} and the specific heat  \cite{Cp} indicate that the second superconducting transition is induced in fields parallel to the ab-plane and that even a slight misalignment of the field suppresses the second superconducting transition. In Fig 4, the $H$-$T$ phase diagrams for $H //$ [110] and $H //$ [100] but with an intentional misalignment of $\Delta \theta = 0.55 ^{\circ}$ are presented, where the feature P$_{2}$ is suppressed. Concomitantly, the unusual temperature dependence of $H_{\rm p}$ seen in Fig. 3 is also suppressed. This fact supports that there is another correlation between the second superconducting transition and the feature P$_{2}$ seen in the ac susceptibility. These two correlations are in principle common to both the [110] and [100] directions, which is consistent with that the second superconducting transition occurs for both $H //$ [110] and $H //$ [100] as evidenced by the specific heat and the thermal conductivity.

\quad Figure 2 additionally shows $H_{\rm c2}$ and $H_{\rm p}$ when the field is along the [110] and [100] directions but with a misalignment of $\Delta \theta = 0.55 ^{\circ}$ (open symbols). The effects of this slight misalignment are rather strong: (1) The in-plane anisotropy of the upper critical field is reduced to 1.6 \%. (2) $H_{\rm 2}$ disappears. (3) $H_{\rm p}$ seems to have a four-fold symmetry, being {\it in-phase} with $H_{\rm c2}$.

\quad As Mao {\it et al}. \cite {Mao} has suggested, the very accurate alignment of the magnetic field to the ab-plane is essential for inducing the second superconducting transition. The enhanced anisotropy of $H_{\rm c2}(\phi)$ and the second superconducting transition both simultaneously appear and disappear when the polar angle $\theta$ is varied across $90^{\circ}$. As the enhanced anisotropy of $H_{\rm c2}$($\phi$) is theoretically expected to accompany the second superconducting transition \cite{theory}, the observed correlation between these supports the application of the Agterberg theory at least in a qualitative fashion.

\quad Next we discuss another aspect that seems to be closely related to the second superconducting transition; the upper critical field $H_{\rm c2}$ immediately above the second superconducting phase is considerably lower than expected values. There are at least two ways of describing this aspect. One is via the dependence of $H_{\rm c2}$ on the polar angle $\theta$. The other is via the shape of the $H$-$T$ phase diagram 
({\it i.e.} comparison with a linear extrapolation of the gradient near $T_{\rm c}$).

\quad First, we show in Fig. 5 the dependences of $H_{\rm c2}$ and $H_{\rm 2}$ on the polar angle $\theta$, with the field inclined towards the [110] and [100] directions. The inset to Fig. 5 illustrates a blow-up of the main panel. Clearly, a misalignment of $\sim$ 0.5$ ^{\circ}$ is large enough to suppress the feature P$_{2}$. Although the measurement temperature 60 mK is considerably lower than the transition temperature 1.46 K, we apply the Ginzburg-Landau anisotropic effective-mass approximation  \cite{tinkham}:
\begin{equation}
H_{{\rm c2}}(\theta) = \frac{H_{\rm c2}(\theta=0)}{\sqrt{\cos^{2}\theta + \mathit{\Gamma}^{-2}\sin^{2}\theta}}.
\end{equation}
Here $\mathit{\Gamma}$ is the square root of the ratio between the effective masses for interplane and in-plane motion, or the ratio between the upper critical fields for in-plane and interplane. Two kinds of fitted curves for $H //$ [110] are shown together with the experimental data in Fig. 5. The solid curve represents a fit of Eq. (1) to the data for $0^{\circ} \le \theta \le 90^{\circ}$ whilst the dashed curve represents a fit for $0^{\circ} \le \theta \le 85^{\circ}$. 
The solid curve as a whole seems to reasonably reproduce the experimental data. The obtained value for $\mathit{\Gamma}$ is 20.1 for $H //$ [110]. The same fitting to the data for $H //$ [100] yields $\mathit{\Gamma}$ = 22.0. These values are in agreement with a previous study  \cite{akima}, in which $\mathit{\Gamma}$ is estimated to be 20. However, it is seen that the solid curve slightly but systematically deviates from the experimental data at low values of $\theta$. On the other hand, the dashed curve (obtained from $0^{\circ} \le \theta \le 85^{\circ}$) in Fig. 5 fits better to low $\theta$ data at the expense of the high $\theta$ region. In fact, the fitting for $0^{\circ} \le \theta \le 85^{\circ}$ yields $H_{\rm c2}(90^{\circ})$ = 1.85 T; the experimentally obtained value for $H_{\rm c2}(90^{\circ})$ being 1.48 T. 
These facts imply the suppression of $H_{\rm c2}$ in a certain range of $\theta$ close to $90^{\circ}$. Since the second superconducting transition occurs for $\theta$ close to $90^{\circ}$ exclusively, this suggests correlation between the second superconducting transition and the suppression of $H_{\rm c2}$.

Second, we use a formula such as the Werthamer-Helfand-Hohenberg (WHH) formula  \cite {WHH} to demonstrate the suppression of $H_{\rm c2}$. Although the WHH formula is intended for orbital depairing in a weak-coupling BCS-type superconductor in the dirty limit, we, for reference, apply the formula
\begin{equation}
\mu_{0}H_{{\rm c2}}(0) = -0.693\mu_{0}\frac{{\rm d}H_{\rm c2}}{{\rm d}T}\bigg|_{T=T_{\rm c}}T_{\rm c},
\end{equation}
to the $H$-$T$ phase diagrams shown in Figs. 3 (a), (b), 4 (a) and (b). Lebed and Hayashi \cite {Lebed} also obtained a WHH-like formula for a quasi two-dimensional {\it p} wave superconductor with an isotropic gap in fields parallel to the ab-plane.
The resultant formula in ref.\cite {Lebed} is identical to Eq.(2) but with the coefficient being $-0.75$ instead of $-0.693$. 
In reality, the application of the WHH formula to the present $H$-$T$ phase diagrams leads to the upper critical field being rather overestimated by a factor of 2-2.5. In other words, the upper critical field at very low temperatures appears to be suppressed compared to that expected from the orbital depairing. This occurs immediately above the second superconducting phase.
This also seemingly suggests that the emergence of the second superconducting transition is closely related to the suppression of low-temperature upper critical fields. However, the $H$-$T$ phase diagram for $\Delta \theta$ = $0.55 ^{\circ}$, at first sight, seems to be a counter example. Possible implications of this will be discussed later.

\quad In contrast, the $H$-$T$ ($H //$ ab-plane) phase diagram established in ref. \cite{synchro} by Yoshida {\it et al}. from ac susceptibility measurements is well explained by the WHH formula. Also the $H$-$T$ ($H //$ c-axis) phase diagram obtained by Mackenzie {\it et al}. from resistive measurements  \cite{andy} fits the same formula very well. As these two works both use samples with $T_{\rm c}$ of $\sim$0.9 K, the discrepancy between the present work and refs. \cite{synchro,andy} may be due to the difference in sample quality. However, it is still unclear whether the second superconducting phase can be induced in samples with $T_{\rm c}$ of $\sim$0.9 K. In ref. \cite{synchro}, the ac susceptibility data taken in magnetic fields parallel to the ab-plane do not show any signs of the second superconducting transition such as P$_{2}$. The possible reasons we suggest are too weak an ac field amplitude (they used ac amplitudes of 3-7 $\mu$T), too large a misalignment (they estimated the actual misalignment to be less than $1.5^{\circ}$) and/or the relatively poor sample quality ($T_{\rm c} \simeq$ 0.9 K).

\quad We now discuss differences in results from various experimental probes, in reference to the second superconducting transition. The greatest difference is the absence of P$_{2}$ in the ac susceptibility for $H //$ [100] whilst the specific heat measurements have observed a clear split at the superconducting transition in fields along the [100] direction \cite{deguchi}. As previously mentioned, this implies that the feature P$_{2}$ in the ac susceptibility does not necessarily represent the second superconducting transition itself albeit P$_{2}$ is closely related to the second superconducting transition. A possible interpretation for the feature P$_{2}$ will be that P$_{2}$ is a consequence of another peak effect due to the second superconducting phase.

\quad In order to support this interpretation, we show the dependence of the ac susceptibility on the ac field amplitude. Fig. 6 shows the real and imaginary parts of the ac susceptibility with different ac amplitudes \cite{degrade}. 
The application of an ac field above $\sim$100 $\mu$T hampers reliable thermometry. The applied dc field direction and the temperature have been chosen to be the [110] direction and 0.43 K, respectively, so that the feature P$_{2}$ is rather clear. 
The feature P$_{2}$ strongly depends on the ac field amplitude, which is particularly clear in the imaginary part, whilst the upper critical field $H_{\rm c2}$ hardly depends on the ac amplitude.
P$_2$ can be seen only above an ac-field amplitude of $\sim$25 $\mu$T and shifts to lower magnetic fields with increasing ac amplitude. It is reported that in the imaginary part of the ac susceptibility of 2{\it H}-NbSe$_{2}$, the peak position is sensitive to the amplitude of the ac field whilst that the upper critical field is hardly affected by the ac amplitude \cite{acfield}. Therefore, the behaviour of P$_{2}$ observed in Fig. 6 strongly suggests that the feature P$_2$ arises from flux pinning rather than from a phase transition itself.

\quad The feature P$_2$ in the real part of the ac susceptibility may be regarded as a hump added to a smoothly varying background. This is strongly suggested by Fig. 6. 
Therefore, the peak feature around $H_2$ can be interpreted as an increase in the real part of the ac susceptibility, {\it i.e.} a decrease in pinning, due to the second superconducting transition.
This stands in contrast to the usual peak effect such as P$_3$ we observe since the usual peak effect is due to an increase in pinning and is observed as a dip in the real part of the ac susceptibility.

\quad Let us here make a remark on the nature of possible phase boundaries in the $H$-$T$ plane. 
As the feature P$_{2}$ seems to mostly occur below the second superconducting transition field \cite{other data}, the feature P$_{2}$ should be mediated by fluctuation, suggestive of the second superconducting transition being second order. Taking into consideration the existence of the bicritical point around $H$ = 1.2 T, $T$ = 0.8 K, the $H_{\rm c2}$ branch above the second superconducting transition line possibly represents a first order transition.

\quad Another significant difference amongst various probes lies in the size of the misalignment that suppresses the signs of the second superconducting transition. The minimum of such a misalignment is $\Delta \theta = 0.5^{\circ}$ for the ac susceptibility and $3^{\circ}$ for the thermal conductivity  \cite{kappa}. This is very likely to be due to the difference in probe. For example, the ac susceptibility measurements involve vortex motion and thus should be considerably more complicated than the thermal conductivity measurements. (When the out-of-plane component of the field becomes large enough for staircase vortices to form, the ac susceptibility will be largely affected. This roughly corresponds to $\Delta \theta = 0.5^{\circ}$ \cite{akima}.) 
It is not straightforward to discuss from existing experimental information how much misalignment of $\Delta \theta$ is large enough to suppress the second superconducting transition. If we closely relate the suppression of the low-temperature $H_{\rm c2}$ to the second superconducting transition, $\Delta \theta = 0.5^{\circ}$ is found to be too small to suppress the second superconducting transition. Also the result of the fitting of Eq. 1 for the range $0^{\circ} \le \theta \le 85^{\circ}$ shown in Fig. 5 perhaps suggests that $H_{\rm c2}$ is suppressed over quite a wide range of $\Delta \theta$.

\section{Summary}
\quad In summary, we have investigated the ac susceptibility of the spin-triplet superconductor Sr$_2$RuO$_4$ down to 60 mK, placing particular importance on multiple superconducting phases. Whilst the present results reproduce the previous results of the ac susceptibility \cite{Mao} very well, there are several discrepancies between the present results and those of recent specific heat and thermal conductivity measurements  \cite{Cp,Mao,kappa,deguchi}. The feature P$_{2}$ seen in the ac susceptibility is very likely to be closely related to the second superconducting transition; however, the previous attribution of P$_{2}$ to the second superconducting transition itself clearly disagrees with the recent studies of the thermal conductivity and the specific heat \cite{kappa,deguchi}. We propose that the feature previously attributed to the second superconducting transition  \cite{Mao} is a consequence of additional involvement of vortex pinning originated from the second superconducting transition. Vortex motion, prominent in the ac susceptibility, appears to be predominantly effective in overshadowing the features of the second superconducting transition.

\begin{acknowledgements}
\quad We thank M. Sigrist, M. A. Tanatar, K. Deguchi and M. Suzuki for invaluable discussions. We also thank E. Ohmichi, Y. Shimojo and N. Kikugawa for technical support during measurements. This work has been in part supported by the Grant-in-Aid for Scientific Research on Priority Area "Novel Quantum Phenomena in Transition Metal Oxides" from the Ministry of Education, Culture, Sports, Science and Technology.
\end{acknowledgements}

\begin{figure} 
\caption{Ac susceptibility ($\chi = \chi^\prime + i \chi^{\prime\prime}$) at various temperatures (a) for $H$ // [110] and (b) for $H$ // [100]. Whilst P$_{1}$, P$_{2}$ and P$_{3}$ are all observed for $H$ // [110], only P$_{1}$ and P$_{3}$ are seen for $H$ // [100]. The ac-field amplitude used is 50 $\mu$T. Traces have been offset for clarity.}
\end{figure}

\begin{figure} 
\caption{Azimuthal-angle $\phi$ dependence of $H_{\rm c2}$, $H_{\rm 2}$ and $H_{\rm p}$ with the magnetic field parallel to the ab-plane ($T =$ 60 mK). $H_{\rm c2}$ is the upper critical field, $H_{\rm 2}$ is related to the second superconducting transition and $H_{\rm p}$ is the second peak due to vortex synchronisation. All of these characteristic fields show clear four-fold symmetry. Smooth curves (solid and dotted lines) through each set of the data points are employed as guides to the eye. The open symbols represent $H_{\rm c2}$ and $H_{\rm p}$ with a slight misalignment of $\Delta \theta = 0.55^{\circ}$. Straight lines (dashed lines) are employed between these symbols.}
\end{figure}

\begin{figure} 
\caption{Field-temperature phase diagrams (a) for $H //$ [110] and (b) for $H //$ [100]. Whilst $H_{\rm p}$ appears for both configurations, $H_{\rm 2}$ appears only for $H //$ [110].}
\end{figure}

\begin{figure} 
\caption{Field-temperature phase diagrams (a) for $H //$ [110] and (b) for $H //$ [100] in misaligned fields ($\Delta \theta = 0.55^{\circ}$).}
\end{figure}

\begin{figure} 
\caption{Polar-angle $\theta$ dependence of $H_{\rm c2}$ and $H_{\rm 2}$ with the field inclined towards the [110] and [100] directions ($T =$ 60 mK). The inset shows a blow-up of the main panel. The solid line and the dashed line represent fits of the Ginzburg-Landau anisotropic effective-mass approximation (Eq.(2)) to the data for $0^{\circ} \le \theta \le 90^{\circ}$ and for $0^{\circ} \le \theta \le 85^{\circ}$, respectively.}
\end{figure}

\begin{figure} 
\caption{Dependence of the ac susceptibility on the ac field amplitude. The data were taken in dc magnetic fields parallel to the [110] direction at 0.43 K. The variation of the appearance of P$_{2}$ with ac field strength indicates that P$_{2}$ involves flux pinning. Traces have been offset for clarity.}
\end{figure}


\begin {references}

\bibitem[*]{maoaddress}
Present address: Physics Department, Tulane University, 2001 Percival Stern, New Orleans, LA 70118.

\bibitem{discovery}
Y. Maeno, H. Hashimoto, K. Yoshida, S. Nishizaki, T. Fujita, J. G. Bednorz, and F. Lichtenberg, Nature (London) {\bf 372}, 532 (1994).

\bibitem{impurity}
A. P. Mackenzie, R. K. W. Haselwimmer, A. W. Tyler, G. G. Lonzarich, Y. Mori, S. Nishizaki and Y. Maeno, Phys. Rev. Lett. {\bf 80}, 161 (1998).

\bibitem{defect}
Z. Q. Mao, Y. Mori, and Y. Maeno, Phys. Rev. B {\bf 60}, 610 (1999).

\bibitem{Rice}
T. M. Rice and M. Sigrist, J. Phys. Condens. Matter {\bf 7}, L643 (1995).

\bibitem{Baskaran}
G. Baskaran, Physica B {\bf 223}-{\bf224}, 490 (1996).

\bibitem{muSR}
G. M. Luke, Y. Fudamoto, K. M. Kojima, M. I. Larkin, J. Merrin, B. Nachumi, Y. J. Uemura, Y. Maeno, Z. Q. Mao, Y. Mori, H. Nakamura, and M. Sigrist, Nature (London) {\bf 394}, 558 (1998).

\bibitem{NMR}
K. Ishida, H. Mukuda, Y. Kitaoka, K. Asayama, Z. Q. Mao, Y. Mori, and Y. Maeno, Nature (London) {\bf 396}, 658 (1998); K. Ishida, H. Mukuda, Y. Kitaoka, Z. Q. Mao, H. Fukazawa, and Y. Maeno, Phys. Rev. B {\bf63}, 060507(R) (2001).

\bibitem{UPt3}
R. A. Fisher, S. Kim, B. F. Woodfield, N. E. Phillips, L. Taillefer, K. Hasselbach, J. Flouquet, A. L. Giorgi, and J. L. Smith  Phys. Rev. Lett. {\bf 62}, 1411 (1989); K. Hasselbach, L. Taillefer, and J. Flouquet, Phys. Rev. Lett. {\bf 63}, 93 (1989).

\bibitem{Cp}
S. NishiZaki, Y. Maeno, and Z. Q. Mao, J. Phys. Soc. Jpn. {\bf 69}, 572 (2000).

\bibitem{NQR}
K. Ishida, H. Mukuda, Y. Kitaoka, Z. Q. Mao, Y. Mori, and Y. Maeno
Phys. Rev. Lett. {\bf 84}, 5387 (2000).

\bibitem{penetration}
I. Bonalde, Brian D. Yanoff, M. B. Salamon, D. J. Van Harlingen, E. M. E. Chia, Z. Q. Mao, and Y. Maeno, Phys. Rev. Lett. {\bf 85}, 4775 (2000).

\bibitem{tanatar}
M. A. Tanatar, M. Suzuki, S. Nagai, Z. Q. Mao, Y. Maeno, and T. Ishiguro, Phys. Rev. Lett. {\bf 86}, 2649 (2001).

\bibitem{izawa}
K. Izawa, H. Takahashi, H. Yamaguchi, Y. Matsuda, M. Suzuki, T. Sasaki, T. Fukase, Y. Yoshida, R. Settai, and Y. Onuki, 
Phys. Rev. Lett. {\bf 86}, 2653 (2001).

\bibitem{polarised}
J. A. Duffy, S. M. Hayden, Y. Maeno, Z. Mao, J. Kulda, and G. J. McIntyre, Phys. Rev. Lett. {\bf 85}, 5412 (2000).

\bibitem{SANS}
P. G. Kealey, T. M. Riseman, E. M. Forgan, L. M. Galvin, A. P. Mackenzie, S. L. Lee, D. McK. Paul, R. Cubitt, D. F. Agterberg, R. Heeb, Z. Q. Mao, and Y. Maeno, Phys. Rev. Lett. {\bf 84}, 6094 (2000).

\bibitem{theory}
D. F. Agterberg, Phys. Rev. Lett. {\bf 80}, 5184 (1998).

\bibitem{Mao}
Z. Q. Mao, Y. Maeno, S. NishiZaki, T. Akima, and T. Ishiguro, Phys. Rev. Lett. {\bf 84}, 991 (2000).

\bibitem{kappa}
M. A. Tanatar, S. Nagai, Z. Q. Mao, Y. Maeno, and T. Ishiguro, Phys. Rev. B{\bf  63}, 064505 (2001).

\bibitem{growth}
Z. Q. Mao, Y. Maeno, and H. Fukazawa, Mater Res. Bull. {\bf 35}, 1813 (2000).

\bibitem{tetragonal}
J. S. Gardner, G. Balakrishnan, D. McK. Paul, and C. Haworth, Physica C {\bf 265}, 251 (1996).

\bibitem{definition}
The characteristic fields $H_{\rm c2}$, $H_{\rm 2}$ and $H_{\rm p}$ have been defined using the real part of the ac susceptibility. These field positions do not always exactly coincide with the peaks in the imaginary part, but show reasonable agreement.

\bibitem{demagnetisation}
Throughout the paper, we do not take into accout the demagnetisation factor. The applied field almost fully penetrates at $H_{\rm c2}$ whilst we will later discuss the dependence of $H_{\rm c2}$ on the angle between the ab-plane and the applied field. When we discuss the characteristic fields $H_{\rm c2}$, $H_{\rm 2}$ and $H_{\rm p}$, all of which are close to $H_{\rm c2}$, the field is parallel to the plate-like sample. The demagnetisation effect is the least important in this field configuration.

\bibitem{synchro}
K. Yoshida, Y. Maeno, S. Nishizaki, and T. Fujita, J. Phys. Soc. Jpn. {\bf 65}, 2220 (1996).

\bibitem{deguchi}
K. Deguchi, M. A. Tanatar, Z. Q. Mao, T. Ishiguro, and Y. Maeno, to be published in J. Phys. Soc. Jpn.

\bibitem{tinkham}
M. Tinkham, {\it ÒIntroduction to SuperconductivityÓ, 2nd edition}, McGraw-Hill, New York, 1996. p. 139.

\bibitem{akima}
T. Akima, S. NishiZaki, and Y. Maeno, J. Phys. Soc. Jpn. {\bf 68}, 694 (1999).

\bibitem{WHH}
N. R. Werthamer, E. Helfand, and P. C. Hohenberg, Phys. Rev. {\bf 147}, 295 (1966).

\bibitem{Lebed}
A. G. Lebed and N. Hayashi, Physica C {\bf 341-348}, 1677 (2000).

\bibitem{andy}
A. P. Mackenzie, S. R. Julian, A. J. Diver, G. G. Lonzarich, N. E. Hussey, Y. Maeno, S. Nishizaki, and T. Fujita, Physica C {\bf 263}, 510 (1996).

\bibitem{degrade}
The data in Fig. 6 are from the same sample as that used for all the other figures in this paper. However, after all the measurements but those for Fig. 6, we found that the $T_{\rm c}$ of the sample was decreased to 1.39 K (by 70 mK). This has probably caused changes of the appearance of the ac susceptibility data and of $H_{\rm c2}$ in Fig. 6.

\bibitem{acfield}
L. A. Angurel, F. Amin, M. Polichetti, J. Aarts, and P. H. Kes, Phys. Rev. B {\bf 56}, 3425 (1997).

\bibitem{other data}
Studies of the thermal conductivity \cite{kappa} and the specific heat  \cite{deguchi} suggest that the difference between the upper critical field and the second superconducting transition field only weakly depends on temperature apart from in the vicinity the bicritical point. The difference at low temperatures is 40-50 mT. If we assume this value for $T =$ 60 mK, $H_{\rm 2}$ (field position of P$_{\rm 2}$) shown in Fig. 2 is mostly below the second superconducting transition field.
\end{references}
\end{document}